\documentclass[a4paper,11pt]{article} 
\usepackage[utf8]{inputenc}
\usepackage{authblk}
\usepackage{fullpage}
\usepackage{graphicx}
\usepackage{xcolor}
\usepackage{algorithm}
\usepackage{algpseudocode}
\usepackage{amssymb}
\usepackage{amsthm}
\usepackage{amsmath}
\usepackage{url}
\usepackage{bbold}
\usepackage[square,numbers,comma,sort&compress]{natbib} 
\usepackage[colorlinks,urlcolor=red,citecolor=red,linkcolor=blue,pdftex, pdfauthor = Lukas Exl]{hyperref}
\usepackage{tabularx}
\usepackage{setspace} 
\newtheorem{thm}{Theorem}
\newtheorem{lem}[thm]{Lemma}
\newtheorem{prop}[thm]{Proposition}
\newtheorem{cor}[thm]{Corollary}
\newtheorem{defi}{Definition}

\newtheorem*{pf}{Proof}
\newcommand{\be}{\begin{equation}}
\newcommand{\ee}{\end{equation}}
\newcommand{\ba}{\begin{array}}
\newcommand{\ea}{\end{array}}
\newcommand{\bea}{\begin{eqnarray}}
\newcommand{\eea}{\end{eqnarray}}
\newcommand{\beas}{\begin{eqnarray*}}
\newcommand{\eeas}{\end{eqnarray*}}

\newcommand{\bh}{{\bf h}}

\newcommand{\bb}{{\bf b}}
\newcommand{\bm}{{\bf m}}
\newcommand{\bn}{{\bf n}}

\title{\textbf{\textsc{A magnetostatic energy formula arising from the $L^2$-orthogonal decomposition of the stray field}}}

\author[1,2,3]{Lukas Exl \thanks{\texttt{lukas.exl@univie.ac.at}}}
\affil[1]{Faculty of Mathematics, University of Vienna, Oskar-Morgenstern-Platz 1, 1090 Wien, Austria.} 
\affil[2]{Institute for Analysis and Scientific Computing, Vienna University of Technology, Wiedner Hauptstra{\ss}e 8-10, 1040, Vienna.} 
\affil[3]{WPI c/o Faculty of Mathematics, University of Vienna, Oskar-Morgenstern-Platz 1, 1090 Wien, Austria.} 
\begin{document}
\maketitle
\textbf{Abstract.} 
A formula for the magnetostatic energy of a finite magnet is proven. In contrast to common approaches, the new energy identity does not rely on evaluation of a nonlocal boundary integral inside the magnet 
or the solution of an equivalent Dirichlet problem. The formula is therefore computationally efficient, which is also shown numerically. 
Algorithms for the simulation of magnetic materials could benefit from incorporating the presented representation of the energy. 
In addition, a natural analogue for the energy via the magnetic induction is given. Proofs are carried out within a setting which is suitable for common discretizations in computational micromagnetics.\\  


\section{Introduction}
The magnetostatic energy of a magnet $\Omega \subset \mathbb{R}^3$ is, up to a constant, given as
\begin{align}\label{energy}
 e_d = -\int_\Omega \boldsymbol{m} \cdot \boldsymbol{h}_s \,dx= \int_\Omega \boldsymbol{m} \cdot \nabla u \,dx,
\end{align}
where the magnetization $\boldsymbol{m}$ is defined in $\Omega$ and zero elsewhere, $\boldsymbol{h}_s = - \nabla u$ is the stray field and $u$ its scalar potential 
satisfying $\Delta u = \nabla \cdot \boldsymbol{m} \,\, \text{in} \,\, \mathbb{R}^3$ \cite{brown1963micromagnetics}. 
In micromagnetic simulations the self-energy \eqref{energy} is known to be the most time-consuming part due to its nonlocal nature \cite{abert_2012_2}.  
Our main contribution is the derivation and proof of the following energy identity 
\begin{align}\label{enform_public}
 e_d = \int_\Omega |\nabla u_0|^2 \,dx+ \frac{1}{4\pi}\int_{\partial \Omega} \int_{\partial \Omega ^\prime} \frac{(\boldsymbol{m}\cdot \boldsymbol{n} - \partial_{\boldsymbol{n}} u_0)(\boldsymbol{m}^\prime\cdot \boldsymbol{n}^\prime - 
 \partial_{\boldsymbol{n}^\prime} u_0^\prime) }{|\boldsymbol{x}-\boldsymbol{x}^\prime|} \,ds_{x^\prime}\,ds_{x},
\end{align}
where 
\begin{align}
 -\Delta u_0 &\,= -\nabla \cdot \boldsymbol{m} \quad \text{in} \,\, \Omega\\
 u_0 &\,= 0 \quad \text{on} \,\,\partial \Omega.
\end{align}

The key observation which leads to the expression \eqref{enform_public} is the $L^2$-orthogonal decomposition of the magnetic field in $\Omega$.  
Note, that neither evaluations of a nonlocal boundary integral inside the domain $\Omega$ nor the solution of an equivalent Dirichlet problem is required to obtain the energy from \eqref{enform_public}. 
This is an interesting observation, since most numerical methods in micromagnetics implement the magnetic self-energy 
in the form \eqref{energy} with first computing the nonlocal field via the convolution with the Green's kernel $G(\boldsymbol{x}) = \tfrac{1}{4\pi} \tfrac{1}{|\boldsymbol{x}|}$  \cite{kronmuller2008handbook,abert_2012_2}  
\begin{align}
 \boldsymbol{h}_s(\boldsymbol{x}) = -\nabla u (\boldsymbol{x}) = \int_{\Omega} \nabla G(\boldsymbol{x}-\boldsymbol{x}^\prime) \,\nabla\cdot \boldsymbol{m}^\prime\,dx^\prime - 
 \int_{\partial \Omega}  \nabla G(\boldsymbol{x}-\boldsymbol{x}^\prime) \, \boldsymbol{m}'\cdot \boldsymbol{n}'\,ds_{x^\prime}
\end{align}

or 
the solution of the PDE 
\begin{align}
   \Delta u = \nabla \cdot \boldsymbol{m} \,\, \text{in} \,\, \mathbb{R}^3                    
\end{align}
with the help of boundary integral operators, which account for the contribution of the field in the external region $\mathbb{R}^3\setminus \overline{\Omega}$. 
In any of these cases, the computation of the energy requires the evaluation of a part of the field by nonlocal convolutions evaluated on the boundary \textit{and} inside the magnet or an additional Dirichlet problem. 
The evaluation of the presented formula, and the analogue to the energy via the vector potential, gets along without these computational tasks. 
Besides the solution of one Dirichlet problem, only the single layer potential has to be evaluated on the boundary, where efficient 
numerical techniques are already available, e.g. \cite{exl2014non}.
This leads to computational advantages if the energy has to be computed directly without relying on the stray field several times in a simulation, as in parts of energy minimization algorithms, e.g., derivative-free 
line search \cite{griewank1986global,oezelt2017transition}, or 
derivative-free methods like simulated annealing \cite{luskin1993numerical,simann}.\\ 
In the following sections we give the main results along with related definitions and illustrate and exemplify the usefulness of the new energy formula numerically. Proofs are given in the final section. %

\section{Main results}

In the following we will use smoothness assumptions which are sufficient to ensure the existence of a unique solution of the stray field problem (compare with Def.~\ref{def:transprob}). This is also a suitable setting for 
common discretization schemes in numerical micromagnetics. However, the results presented here are certainly compatible with higher order regularity assumptions. 
For a rigorous presentation of the following definitions we refer to the literature \cite{steinbach2007numerical,sauter2011boundary,mclean2000strongly}.
Let $\Omega \subset \mathbb{R}^3$ be a bounded Lipschitz-domain with boundary $\Gamma:=\partial \Omega$. We denote the exterior domain with $\Omega^{ext}:=\mathbb{R}^3\setminus \overline{\Omega}$. 
We will make use of the Sobolev spaces 
$H^1(\Omega) := W^{1,2}(\Omega) = \{u\in L^2(\Omega):\, \text{weak derivatives}\,\, \partial_q u \in L^2(\Omega),\, q=1,2,3\}$ and 
$H^1_{loc}(\Omega^{ext}):= \{u\in H^1(C):\, C \subset \Omega^{ext} \,\, \text{compact}\}$.  For the definition of the Sobolev spaces on manifolds, in particular $H^{1/2}(\Gamma)$ and its dual space $H^{-1/2}(\Gamma)$, we refer to 
the literature. We denote $H^1_0(\Omega) := \{u \in H^1(\Omega):\, \gamma_0 u= 0 \,\,\text{on}\,\, \Gamma\}$, 
where $\gamma_0: H^1(\Omega) \rightarrow H^{1/2}(\Gamma)$ is the trace. Further, we use the short notation $\langle u,v\rangle_\Gamma = (u,v)_{L^2(\Gamma)}$. 
The conormal derivative $\gamma_1^{int} u \in H^{-1/2}(\Gamma)$ is defined as the solution of the variational problem 
$\langle\gamma_1^{int} u ,\gamma_0^{int} v \rangle_\Gamma =  (\nabla u, \nabla v)_{L^2(\Omega)} - \langle f, v\rangle$ for all $v \in H^1(\Omega)$, where $u \in H^1(\Omega)$ 
satisfies $\Delta u = f,\,\,f \in H^1(\Omega)^\ast,$ in the sense of distributions; the exterior conormal derivative 
$\gamma_1^{ext}: H^1_{loc}(\Omega^{ext}) \rightarrow H^{-1/2}(\Gamma)$ is defined accordingly. 
The expression on the boundary $\boldsymbol{m}\cdot \boldsymbol{n}$, where $\boldsymbol{n}$ is the outer normal, is defined by the bounded linear map 
$\gamma_{\boldsymbol{n}}:\, \big(H^1(\Omega)\big)^3 \rightarrow H^{-1/2}(\Gamma), \, \gamma_{\boldsymbol{n}}(\boldsymbol{m})=\boldsymbol{m} \cdot \boldsymbol{n}$ \cite{girault2012finite}. 

The scalar potential can be characterized via the following transmission problem.
\begin{defi}[Transmission problem]\label{def:transprob}
 Let $\boldsymbol{m}\in \big(H^1(\Omega)\big)^3$. Then the scalar potential $u = (u^{int},u^{ext}) \in H^1(\Omega) \times H^1_{loc}(\Omega^{ext})$ is the unique solution \cite{carstensen1995adaptive} of 
 \begin{eqnarray}\label{transprob}
  \begin{aligned}
   -\Delta u^{int} & \,= - \nabla \cdot \boldsymbol{m} \quad \text{in} \,\, \Omega\\
   \gamma_0^{int}u^{int} &\, =  \gamma_0^{ext}u^{ext}  \quad \text{on} \,\, \Gamma\\
   \gamma_1^{int} u^{int} &\, =   \gamma_1^{ext} u^{ext} + \boldsymbol{m} \cdot \boldsymbol{n}  \quad \text{on} \,\, \Gamma\\
   -\Delta u^{ext} & \,= 0 \quad \text{in} \,\, \Omega^{ext}\\
   u^{ext} &\,= \mathcal{O}(|\boldsymbol{x}|^{-1}) \quad |\boldsymbol{x}| \rightarrow \infty,
  \end{aligned}
 \end{eqnarray}
 where $\boldsymbol{n}$ denotes the outer normal vector. 
 The stray field is $\boldsymbol{h}_s = - \nabla u$.
 \end{defi}

A solution of \eqref{transprob} can be represented with the help of the \textit{single layer potential}.
\begin{defi}[Single layer potential \cite{steinbach2007numerical}]
 The single layer potential $\widetilde{\mathcal{V}}: H^{-1/2}(\Gamma) \rightarrow H^1_{loc}(\mathbb{R}^3)$ is 
 \begin{align}
  (\widetilde{\mathcal{V}}\phi)(\boldsymbol{x}) := \int_{\Gamma} G(\boldsymbol{y}-\boldsymbol{x})\phi(\boldsymbol{y})ds_{y},
 \end{align}
 where $G(\boldsymbol{x}) = \tfrac{1}{4\pi} \tfrac{1}{|\boldsymbol{x}|}$ is the Green's function of the Laplacian in $\mathbb{R}^3$.
\end{defi}
There holds \cite{steinbach2007numerical}
\begin{align}
 \Delta \widetilde{\mathcal{V}}\phi & \,= 0 \quad \text{in} \,\, \Omega \cup \Omega^{ext}.
\end{align}
The linear operator $\widetilde{\mathcal{V}}$ is continuous, while the conormal derivative jumps on $\Gamma$, i.e.,
\begin{align}
 \gamma_0^{int}\widetilde{\mathcal{V}}\phi -  \gamma_0^{ext}\widetilde{\mathcal{V}}\phi &\,= 0\\
 \gamma_1^{int}\widetilde{\mathcal{V}}\phi -  \gamma_1^{ext}\widetilde{\mathcal{V}}\phi &\,= \phi.
\end{align}
There holds $\widetilde{\mathcal{V}}(\boldsymbol{x}) = \mathcal{O}(|\boldsymbol{x}|^{-1})$ as $|\boldsymbol{x}| \rightarrow \infty$.
We define the trace of the single layer potential $\mathcal{V} := \gamma_0 \widetilde{\mathcal{V}}: H^{-1/2}(\Gamma) \rightarrow H^{1/2}(\Gamma)$, which is then a bounded linear operator.\\

The following can be verified, by using the above mentioned properties of the single layer potential.
\begin{prop}[\cite{garcia2006adaptive}]\label{sol}
The solution to \eqref{transprob} is given by 
$u = \big(u_0 + \widetilde{\mathcal{V}}(\boldsymbol{m}\cdot \boldsymbol{n} - \gamma_1^{int} u_0), \widetilde{\mathcal{V}}(\boldsymbol{m}\cdot \boldsymbol{n} - \gamma_1^{int} u_0)\big)$, where
\begin{eqnarray}\label{u_0}
\begin{aligned}
  -\Delta u_0 &\,= -\nabla \cdot \boldsymbol{m} \quad \text{in} \,\, \Omega\\
  \gamma_0^{int} u_0 &\,= 0 \quad \text{on} \,\,\Gamma.
\end{aligned}
\end{eqnarray}

Hence, the stray field in whole space is given as 
\begin{align}\label{solution}
\boldsymbol{h}_s = \boldsymbol{h}_{s,0} + \boldsymbol{h}_{s,1}:=-\nabla u_0 - \nabla \widetilde{\mathcal{V}}(\boldsymbol{m}\cdot \boldsymbol{n} - \gamma_1^{int} u_0),                                  
\end{align}
where $u_0$ satisfies \eqref{u_0} in $\Omega$ and is extended with zero to the exterior domain. \hfill$\square$
\end{prop}

The main result is the following energy identity corresponding to formula \eqref{enform_public}.

\begin{thm}[Energy formula]\label{Thm.1}
 Let $\boldsymbol{h}_{s,0} = -\nabla u_0$ with $u_0 \in H^1_0(\Omega)$ fulfilling the Dirichlet problem \eqref{u_0}. Then the magnetostatic energy 
 is given by
 \begin{align}\label{energyform}
  e_d = -(\boldsymbol{h}_s,\boldsymbol{m})_{(L^2(\Omega))^3} = \|\boldsymbol{h}_{s,0}\|_{(L^2(\Omega))^3}^2 + \langle (\boldsymbol{m}\cdot \boldsymbol{n}-\gamma_1^{int}u_0,\mathcal{V}\big(\boldsymbol{m}\cdot \boldsymbol{n}-\gamma_1^{int}u_0\big)  \rangle_\Gamma.
 \end{align}\hfill $\square$
\end{thm}

An analog energy formula can be derived via the magnetic induction and its vector potential.
\begin{defi}[Transmission problem for the vector potential]\label{def:transprob_vec}
 Let $\boldsymbol{m}\in \big(H^1(\Omega)\big)^3$. Then the vector potential $\boldsymbol{A} = (\boldsymbol{A}^{int},\boldsymbol{A}^{ext}) \in \big(H^1(\Omega)\big)^3 \times \big(H^1_{loc}(\Omega^{ext})\big)^3$ is the unique solution of
\begin{subequations}\label{transp_vec}
\begin{alignat}{3}
 &\Delta \boldsymbol{A}^{int}=-\nabla\times \bm&&\text{ in }\Omega,\\
 &\Delta \boldsymbol{A}^{ext}= \mathbf{0}&&\text{ in }\Omega^{ext},\\
 &\gamma_0^{int}\boldsymbol{A}^{int}=\gamma_0^{ext}\boldsymbol{A}^{ext}&&\text{ on }\Gamma,\\
 &\gamma_1^{int}\boldsymbol{A}^{int}=\gamma_1^{ext}\boldsymbol{A}^{ext}+\bm\times \bn&&\text{ on }\Gamma,\\
 &\boldsymbol{A}_j^{ext}=O(|x|^{-1}),\,\,j=1\hdots3&&\,\,\text{ in }\Omega^{ext},
 \end{alignat}
 \end{subequations} where $\boldsymbol{n}$ denotes the outer normal vector. The trace operators and conormal derivatives are applied component-wisely.
\end{defi}
The analogue to Thm.~\ref{Thm.1} is given next.
\begin{thm}[Energy formula]\label{Thm.2}
 Let $\boldsymbol{b}_{0}^\prime = \nabla \times \boldsymbol{A}_0$ with $\boldsymbol{A}_0 \in (H^1_0(\Omega))^3$ fulfilling the Dirichlet problem \eqref{A_0}. Then the magnetostatic energy 
 is given by
 \begin{align}\label{energyform2}
  e_d = \|\boldsymbol{m}\|_{(L^2(\Omega))^3}^2- \big(\|\nabla \boldsymbol{A}_0\|_{L^2(\Omega)^{3\times 3}}^2 + \langle \boldsymbol{m} \times \boldsymbol{n} - \gamma_1^{int} \boldsymbol{A}_{0},
  \mathcal{V}\big( \boldsymbol{m} \times \boldsymbol{n} - \gamma_1^{int} \boldsymbol{A}_{0}\big)  \rangle_{\Gamma^3}.
 \end{align}\hfill $\square$
\end{thm}

\section{Numerical validation}
In the following we will give numerical results that demonstrate that the presented formulation in Eqn.~\eqref{enform_public} leads to properly calculated magnetostatic energy. 
We perform our computations on finite element grids with P1-elements and compare with FEM/BEM for \eqref{transprob} via the representation in Prop.~\ref{sol} using a mass-lumped stray field for the energy computation \cite{schreflFEM} 
via Eqn.~\eqref{energy}. This approach is common in micromagnetics \cite{garcia2006adaptive,exl2014non} and requires the solution of two Dirichlet problems, the one in Eqn.~\eqref{u_0} and a second one with zero right hand side and Dirichlet data obtained from the evaluation of the single layer potential on the boundary nodes.
In contrast to that, the new energy formula does not require solving the latter.     
For the efficient computation of the single layer potential on the boundary nodes in (quasi) linear time (and linear memory consumption) the NUFFT method in \cite{exl2014non} is used. In all computations 
the normal component of the magnetization in the source term of the single layer potential was projected onto the space of piecewise constant functions ($L^2$-projection). Likewise, the $L^2$-projection is used for a piecewise constant 
approximation of the $\mathcal{V}$-surface values in the trace product, cf. Eqn. \eqref{energyform}. In the FEM/BEM approach a nodal interpolation of the $\mathcal{V}$-surface values for the Dirichlet data is used for the sake of effectiveness.    
The Dirichlet problems are solved with an ILU preconditioned conjugate gradient method. Computations were performed on the Vienna Scientific Cluster 3 (VSC3). Mesh generation was done with help of NETGEN \cite{schoberl1997netgen}.  
Results for the uniformly magnetized unit cube are given in Tab.~\ref{tab1}, for a uniformly magnetized sphere in Tab.~\ref{tab2} and for some random configuration in the cube in Tab.~\ref{tab3}. 
For the cases of a uniformly magnetized sphere or cube the results are also compared with the analytical values. 
All test cases show accurate results for the new formula and a gain in efficiency from $12-25\%$ relative to the FEM/BEM approach. 
This mostly amounts to the cost of solving one additional inhomogeneous Dirichlet problem in the FEM/BEM approach, which is particularly apparent for bulk material, but gets less relevant for, e.g., thin film geometries.
\begin{table}
\tabcolsep 2pt 
\caption{Errors and timings for uniformly magnetized unit cube with analytical value $1/6\,[\mu_0 M_s^2]$ (including a factor $1/2$ in formula \eqref{energy}, 
$\mu_0$ denotes the vacuum permeability and $M_s$ the saturation magnetization). 
Mesh data: number of nodes ($\#$nodes), number of surface nodes ($\#$snodes), number of surface triangles 
($\#$stri) and number of tetrahedral elements ($\#$tets). Energies: Energy value computed with formula~\eqref{enform_public} ($E_{new}$) and with FEM/BEM ($E_{FB}$). 
Deviations/Errors: Relative deviation of energies computed with new formula~\eqref{enform_public} and FEM/BEM (dNewFB), reference value and FEM/BEM (dRefFB) and reference value and new formula (dRefNew).   
Timings: Computation time via FEM/BEM ($t_{FB}$), computation time for formula \eqref{enform_public} ($t_{new}$) and gain (in $\%$).} \label{tab1}
\begin{center}
\begin{tabular}{cccccccccccc}
$\#$nodes    & $\#$snodes & $\#$stri &  $\#$tets & $E_{new}$ & $E_{FB}$ & dNewFB & dRefFB & dRefNew  & $t_{FB}\,[s]$ & $t_{new}\,[s]$ &  gain \\ \hline\hline
2744 & 1016 & 2028 & 13182 & 1.652E-01 & 1.652E-01 & 0        & 8.32E-03 & 8.32E-03 & 0.19 & 0.17 & 12$\%$\\ \hline  
9261 & 2402 & 4800 & 48000 & 1.659E-01 & 1.659E-01 & 0        & 4.05E-03 & 4.05E-03 & 0.31 & 0.27 & 12$\%$\\ \hline
19683 & 4058 & 8112& 105456& 1.662E-01 & 1.662E-01 & 5.96E-05 & 2.49E-03 & 2.43E-03 & 0.47 & 0.39 & 17$\%$\\ \hline 
68921 & 9602 &19200& 384000& 1.665E-01 & 1.665E-01 & 1.38e-05 & 1.13E-03 & 1.11E-03 & 1.28 & 0.97 & 25$\%$\\ \hline 
\end{tabular}
\end{center}
\end{table}
\begin{table}
\tabcolsep 2pt 
\caption{Errors and timings for uniformly magnetized sphere with radius $0.5$ with analytical value $8.727$-$E02\,[\mu_0 M_s^2]$ (including a factor $1/2$ in formula \eqref{energy}, 
$\mu_0$ denotes the vacuum permeability and $M_s$ the saturation magnetization). 
Mesh data: number of nodes ($\#$nodes), number of surface nodes ($\#$snodes), number of surface triangles 
($\#$stri) and number of tetrahedral elements ($\#$tets). Energies: Energy value computed with formula~\eqref{enform_public} ($E_{new}$) and with FEM/BEM ($E_{FB}$). 
Deviations/Errors: Relative deviation of energies computed with new formula~\eqref{enform_public} and FEM/BEM (dNewFB), reference value and FEM/BEM (dRefFB) and reference value and new formula (dRefNew).  
Timings: Computation time via FEM/BEM ($t_{FB}$), computation time for formula \eqref{enform_public} ($t_{new}$) and gain (in $\%$).} \label{tab2}
\begin{center}
\begin{tabular}{cccccccccccc}
$\#$nodes    & $\#$snodes & $\#$stri &  $\#$tets & $E_{new}$ & $E_{FB}$ & dNewFB & dRefFB & dRefNew  & $t_{FB}\,[s]$ & $t_{new}\,[s]$ &  gain \\ \hline\hline
2824 & 891 & 1778 & 14156 & 8.642E-02 & 8.642E-02 & 1.35e-05 & 9.60E-03 & 9.61E-03 & 0.19 & 0.16 & 14$\%$\\ \hline 
4749 & 1295 & 2586 & 24474 & 8.669E-02 & 8.669E-02 & 8.15E-06 & 6.64E-03 & 6.63E-03 & 0.21 & 0.18 & 15$\%$\\ \hline 
9084 & 2031 & 4058 & 48111 & 8.689E-02 & 8.689E-02 & 1.79E-08 & 4.23E-03 & 4.23E-03 & 0.34 & 0.29 & 15$\%$\\ \hline 
13837& 4098 & 8192 & 70144 & 8.709E-02 & 8.709E-02 & 1.79e-07 & 2.07E-03 & 2.07E-03 & 0.58 & 0.49 & 15 $\%$\\ \hline
\end{tabular}
\end{center}
\end{table}
\begin{table}
\tabcolsep 2pt 
\caption{Errors and timings for randomly magnetized unit cube, deviating from uniform magnetization at nodes by normally distributed polar angle with zero mean and standard deviation of $20$ degrees. 
Mesh data: number of nodes ($\#$nodes), number of surface nodes ($\#$snodes), number of surface triangles 
($\#$stri) and number of tetrahedral elements ($\#$tets). Energies: Energy value computed with formula~\eqref{enform_public} ($E_{new}$) and with FEM/BEM ($E_{FB}$). 
Deviations: Relative deviation of energies computed with new formula~\eqref{enform_public} and FEM/BEM (dNewFB).   
Timings: Computation time via FEM/BEM ($t_{FB}$), computation time for formula \eqref{enform_public} ($t_{new}$) and gain (in $\%$).} \label{tab3}
\begin{center}
\begin{tabular}{cccccccccc}
$\#$nodes    & $\#$snodes & $\#$stri &  $\#$tets & $E_{new}$ & $E_{FB}$ & dNewFB &  $t_{FB}\,[s]$ & $t_{new}\,[s]$ & gain \\ \hline\hline
2744 & 1016 & 2028 & 13182 & 1.491E-01 & 1.491E-01 & 1.82E-04  &  0.19 & 0.17 & 12$\%$\\ \hline  
9261 & 2402 & 4800 & 48000 & 1.502E-01 & 1.503E-01 & 1.00E-03 & 0.30 & 0.26 & 12$\%$\\ \hline
19683 & 4058 & 8112 & 105456 & 1.503E-01 & 1.506E-01 & 1.90E-03 & 0.46 & 0.39 & 15$\%$\\ \hline
68921 & 9602 & 19200 & 384000 & 1.505E-01 & 1.507E-01 & 1.97E-03 & 1.57 & 1.16 & 26$\%$\\ \hline
\end{tabular}
\end{center}
\end{table}

\section{Proofs of the energy identities}
There holds the following $L^2$-orthogonality. 
\begin{lem}[$L^2$-orthogonality]\label{l2orth}
 Let $u_0 \in H^1_0(\Omega)$ be the solution to the Dirichlet problem \eqref{u_0} and $\boldsymbol{h}_0 = -\nabla u_0 \in (L^2(\Omega))^3$ the corresponding field. Let further 
 $\boldsymbol{h}_\Delta \in (L^2(\Omega))^3$ be a Laplace field, i.e., $\nabla\cdot \boldsymbol{h}_\Delta = 0$ in $\Omega$. Then 
 \begin{align}
  (\boldsymbol{h}_{0}, \boldsymbol{h}_{\Delta} )_{(L^2(\Omega))^3} := \int_\Omega \boldsymbol{h}_0\cdot \boldsymbol{h}_\Delta \,dx= 0.
 \end{align}
\end{lem}
\begin{pf}
 By partial integration on gets 
 \begin{align}
 (-\nabla u_0, \boldsymbol{h}_{\Delta} )_{(L^2(\Omega))^3} = (u_0, \nabla \cdot \boldsymbol{h}_{\Delta} )_{L^2(\Omega)} - \langle \gamma_0^{int} u_0,\boldsymbol{h}_\Delta\cdot \boldsymbol{n}\rangle_{\Gamma}.
 \end{align}
 Both terms on the r.h.s. are zero.
\hfill $\square$
\end{pf}
For the components in Eqn.~\eqref{solution} we conclude.
\begin{cor}\label{cor1}
For the solution to \eqref{transprob} given by 
$u = u_0 + u_1$ with $u_1:=\widetilde{\mathcal{V}}(\boldsymbol{m}\cdot \boldsymbol{n} - \gamma_1^{int} u_0)$ and $u_0$ from \eqref{u_0} 
the corresponding fields $\boldsymbol{h}_{s,0} = -\nabla u_0$ and $\boldsymbol{h}_{s,1}=-\nabla u_1$ are $L^2$-orthogonal, i.e., 
 \begin{align}
  (\boldsymbol{h}_{s,0}, \boldsymbol{h}_{s,1} )_{(L^2(\mathbb{R}^3))^3} =  (\boldsymbol{h}_{s,0}, \boldsymbol{h}_{s,1} )_{(L^2(\Omega))^3} = 0. 
 \end{align}
\hfill $\square$
\end{cor}
The formula \eqref{enform_public} (Theorem~\ref{Thm.1}) follows with help of Cor.~\ref{cor1}.\\

\noindent\textbf{Proof of Thm.~\ref{Thm.1} \quad}
 By Thm.~2.7.7 in \cite{sauter2011boundary} we get from \eqref{u_0} with $v \in H^1(\Omega)$
 \begin{align}
  \langle\gamma_1^{int}u_0 , \gamma_0^{int}v\rangle_{\Gamma} = (\nabla u_0, \nabla v)_{(L^2(\Omega))^3} + (\nabla \cdot \boldsymbol{m},v)_{L^2(\Omega)}.
 \end{align}
Now, applying Green's first identity gives
 \begin{align}
  \langle\gamma_1^{int}u_0 , \gamma_0^{int}v\rangle_{\Gamma} = (\nabla u_0, \nabla v)_{(L^2(\Omega))^3} - (\boldsymbol{m},\nabla v)_{(L^2(\Omega))^3} 
  + \langle\boldsymbol{m}\cdot \boldsymbol{n}, v\rangle_{\Gamma}.
 \end{align}
 Rearranging terms gives
 \begin{align}
  \langle\gamma_1^{int}u_0 - \boldsymbol{m}\cdot \boldsymbol{n}, \gamma_0^{int}v\rangle_{\Gamma} = (\nabla u_0, \nabla v)_{(L^2(\Omega))^3} - (\boldsymbol{m},\nabla v)_{(L^2(\Omega))^3}.
 \end{align}
We insert $v:=u=u_0 + u_1 \in H^1(\Omega)$ with $u_1:=\widetilde{\mathcal{V}}(\boldsymbol{m}\cdot \boldsymbol{n} - \gamma_1^{int} u_0)$ and $u_0$ from \eqref{u_0} and 
use the zero-boundary condition of $u_0$ and the orthogonality from Cor.~\ref{cor1}. This yields 
\begin{align}\label{result}
 \langle\gamma_1^{int}u_0 - \boldsymbol{m}\cdot \boldsymbol{n}, \gamma_0^{int}u_1\rangle_{\Gamma} = \|\nabla u_0\|_{(L^2(\Omega))^3}^2 - (\boldsymbol{m},\nabla u)_{(L^2(\Omega))^3}, 
\end{align}
which immediately gives formula \eqref{energyform}.\hfill$\square$\\

There is an alternative derivation of \eqref{energyform} via the magnetic induction.\\

\noindent\textbf{Alternative proof of Thm.~\ref{Thm.1} \quad}
  We use the fundamental Helmholtz decomposition (omitting constants) \cite{brown1963micromagnetics,aharoni2000introduction}
 \begin{align}\label{helmholtz}
  \boldsymbol{m} = \boldsymbol{b}^\prime - \boldsymbol{h}_s \quad \text{in}\,\,\mathbb{R}^3,
 \end{align}
 where $\boldsymbol{b}'$ is the divergence-free magnetic induction (not including the part corresponding to the external field). The fields $\boldsymbol{h}_s,\boldsymbol{b}^\prime \in (L^2(\mathbb{R}^3))^3$ are $L^2$- orthogonal \cite{brown1963micromagnetics}. 
 A consequence of this is that the energy can be written as
\begin{align}\label{energy_wholespace}
 e_d = -\int_\Omega \boldsymbol{m} \cdot \boldsymbol{h}_s \,dx= \int_{\mathbb{R}^3} |\boldsymbol{h}_s|^2 \,dx.
\end{align}
 
We can now use the representation $\boldsymbol{h}_s = \boldsymbol{h}_{s,0} + \boldsymbol{h}_{s,1}:=-\nabla u_0 - \nabla u_1,\,\,u_1=\widetilde{\mathcal{V}}(\boldsymbol{m}\cdot \boldsymbol{n} - \gamma_1^{int} u_0)$ for the stray field in 
whole space, where $u_0$ is extended with zero in the exterior domain $\Omega^{ext}$. Cor.~\ref{cor1} yields
\begin{align}
 e_d = \int_{\Omega} |\boldsymbol{h}_{s,0}|^2 \,dx + \int_{\mathbb{R}^3} |\boldsymbol{h}_{s,1}|^2 \,dx =  \|\nabla u_0\|_{(L^2(\Omega))^3}^2 + \int_{\Omega} |\boldsymbol{h}_{s,1}|^2 \,dx + \int_{\Omega^{ext}} |\boldsymbol{h}_{s,1}|^2 \,dx.
\end{align}
By applying Green's first formula we get \cite[Ch.~6.6.1]{steinbach2007numerical}
\begin{align}
 \int_{\Omega} |\boldsymbol{h}_{s,1}|^2 \,dx &\,= \langle \gamma_1^{int} u_1, \gamma_0 u_1 \rangle_\Gamma\\
 \int_{\Omega^{ext}} |\boldsymbol{h}_{s,1}|^2 \,dx &\,= \langle -\gamma_1^{ext} u_1, \gamma_0 u_1 \rangle_\Gamma
\end{align}
The jump of the conormal derivative $\gamma_1^{int}u_1 -  \gamma_1^{ext}u_1 = \boldsymbol{m}\cdot\boldsymbol{n} - \gamma_1^{int} u_0$ leads to the result.
\hfill$\square$\\

We derive and proof now the analog Thm.~\ref{Thm.2}.\\

\noindent\textbf{Proof of Thm.~\ref{Thm.2} \quad} 
Maxwell's equations for magnetostatics read
\begin{align}
 \nabla \cdot \bb \,&\,= 0,\\
 \nabla \times \bh \,&\, = \mathbf{j},
\end{align}
where $\mathbf{j}$ is the current density.
Excluding the divergence free part $\boldsymbol{h}_{ext}:\, \nabla \times \boldsymbol{h}_{ext} = \mathbf{j}$ from $\boldsymbol{h} = \boldsymbol{h}_{s} + \boldsymbol{h}_{ext}$ 
gives together with $\bm = \bb^\prime - \bh_s$ 
\begin{align}
\nabla \cdot \bb^\prime \,&\,= 0,\\
 \nabla \times \bb^\prime \,&\, = \nabla \times \bm. 
\end{align}
By introducing a vector potential $\bb^\prime = \nabla \times \boldsymbol{A}$ with \textit{gauge condition} $\nabla\cdot \boldsymbol{A} = 0$ \cite{jackson1999classical}, we get
\begin{align}
 \nabla \times (\nabla \times \boldsymbol{A})  = \nabla(\nabla\cdot \boldsymbol{A}) - \Delta \boldsymbol{A} \,&\, = \nabla \times \bm,
\end{align}
and hence 
\begin{align}
 \Delta \boldsymbol{A} \,&\, = -\nabla \times \bm \quad \text{in}\,\,\mathbb{R}^3.
\end{align}
For a finite magnet $\Omega \subset \mathbb{R}^3$ this gives the boundary conditions \cite{brown1962magnetostatic}
\begin{align}
 \gamma_0 \boldsymbol{A}^{ext} - \gamma_0 \boldsymbol{A}^{int} \,&\,= \mathbf{0}, \\  
 \gamma_1 \boldsymbol{A}^{ext} - \gamma_1 \boldsymbol{A}^{int} \,&\,= -\bm \times \bn.
\end{align}
The analogy of the transmission problem in Def.~\ref{def:transprob_vec} to that of Def.~\ref{def:transprob} gives rise to the representation of the solution for the magnetic induction
\begin{align}
 \boldsymbol{b}^\prime = \boldsymbol{b}_{0}^\prime + \boldsymbol{b}_{1}^\prime:= \nabla \times \boldsymbol{A}_0 + \nabla \times \boldsymbol{A}_1\,\,\text{with}\,\, 
 \boldsymbol{A}_1 = \widetilde{\mathcal{V}}(\boldsymbol{m}\times \boldsymbol{n} - \gamma_1^{int} \boldsymbol{A}_0),
\end{align}
where $\widetilde{\mathcal{V}}$ is applied component-wisely and $\boldsymbol{A}_0 \in \big(H_0^1(\Omega)\big)^3$ fulfilling
\begin{eqnarray}\label{A_0}
\begin{aligned}
  \Delta \boldsymbol{A}_0 &\,= -\nabla \times \boldsymbol{m} \quad \text{in} \,\, \Omega\\
  \gamma_0^{int}\boldsymbol{A}_0 &\,= \boldsymbol{0} \quad \text{on} \,\,\Gamma.
\end{aligned}
\end{eqnarray}
The magnetostatic energy can be expressed in terms of the vector potential by using the decomposition \eqref{helmholtz}  
\begin{align}
  e_d = -\int_\Omega \boldsymbol{m} \cdot \boldsymbol{h}_s \,dx = \int_\Omega |\boldsymbol{m}|^2 \,dx -  \int_\Omega \boldsymbol{m} \cdot \boldsymbol{b}^\prime \,dx. 
\end{align}
The rest of the proof of Thm.~\ref{Thm.2} goes along the same lines as that of Thm.~\ref{Thm.1} but using the orthogonality
 \begin{align}
  (\nabla \boldsymbol{A}_0,\nabla \boldsymbol{A}_1)_{L^2(\Omega)^{3\times 3}} = 0, 
 \end{align}
and the Green's identity (e.g. \cite{sheen1992generalized}) for $\boldsymbol{V} \in \big(H^1(\Omega)\big)^3$ 
\begin{align}
  (\nabla \times \boldsymbol{m},\boldsymbol{V})_{L^2(\Omega)^{3}} = (\nabla \times \boldsymbol{V},\boldsymbol{m})_{L^2(\Omega)^{3}} - 
  \langle \boldsymbol{m}\times \boldsymbol{n},\boldsymbol{V} \rangle_{\Gamma^3}.
\end{align}
\hfill$\square$

\section{Conclusion}
We have derived, proven and numerically validated a new formula for the stray field energy. Computation of the energy only requires the solution of a Dirichlet problem and the evaluation of the 
single layer potential on the boundary. The efficiency is illustrated by means of numerical examples indicating a gain up to $25\%$ compared to a FEM/BEM approach with quasi-optimally scaling BEM part.
The setting was chosen to be suitable for common discretizations in numerical micromagnetics. The presented energy formula has an analogue via the magnetic induction. 
Numerical software for simulation of magnetic materials could benefit in terms of efficiency from incorporating the presented formulas. 

\section*{Acknowledgments}
Financial support by the Austrian Science Fund (FWF) via the SFB "ViCoM" under grant No. F41, via the SFB "Complexity in PDEs" under grant F65 and the project "ROAM" under grant No. 31140 is acknowledged. 
The computations were achieved by using the Vienna Scientific Cluster (VSC) via the funded project No. 71140.

\bibliographystyle{abbrv} 
\bibliography{bibref}
\end{document}